\documentclass[a4paper,twocolumn,11pt,unpublished]{quantumarticle}
\pdfoutput=1
\usepackage[utf8]{inputenc}
\usepackage[english]{babel}
\usepackage[T1]{fontenc}
\usepackage{amsmath}
\usepackage{amssymb}
\usepackage{amsthm}
\usepackage{booktabs}
\usepackage{graphicx}
\usepackage{array}
\usepackage{xcolor}
\usepackage{listings}
\usepackage[numbers,sort&compress]{natbib}
\usepackage{hyperref}

\lstdefinestyle{paperpseudo}{
  basicstyle=\ttfamily\scriptsize,
  columns=fullflexible,
  keepspaces=true,
  frame=single,
  rulecolor=\color{black!25},
  xleftmargin=1em,
  xrightmargin=1em,
  breaklines=true,
  showstringspaces=false
}

\hypersetup{
  colorlinks=true,
  linkcolor=blue!55!black,
  citecolor=blue!55!black,
  urlcolor=blue!55!black
}

\newcommand{\qaspa}{QASPA-MaxCut}
\newcommand{\qaoa}{QAOA}

\newcommand{\ketplus}{|+\rangle}

\makeatletter
\newcommand{\listingneedspace}[1]{%
  \par\begingroup
  \dimen@=#1\relax
  \dimen@ii=\pagegoal
  \advance\dimen@ii by -\pagetotal
  \ifdim\dimen@>\dimen@ii
    \break
  \fi
  \endgroup
}
\def\@printtitle{%
  \textcolor{quantumviolet}{\@printtitletextwithappropriatefontsize\unskip}%
}
\makeatother

\begin{document}

\title{QASPA-MaxCut on Unweighted Graphs: Edge Encoding and QAOA Restarts}

\author{Dmytro Sapozhnyk}
\maketitle

\hypersetup{
  pdftitle={QASPA-MaxCut on Unweighted Graphs: Edge Encoding and QAOA Restarts},
  pdfauthor={Dmytro Sapozhnyk}
}

\begin{abstract}
Max-Cut is a common testbed for near-term quantum optimization because it has a
simple Ising Hamiltonian, established classical baselines, and a parameterized
quantum circuit whose performance depends on the outer classical optimizer.
This paper develops \qaspa, a shift-phase circuit for Max-Cut that
encodes graph edges through controlled $R_y$ rotations of the form
$\mathrm{CX}_{ij}R_y(\theta_{ij})\mathrm{CX}_{ij}$.  We derive the exact
two-qubit algebra of the edge block, trace an unweighted four-vertex example,
and compare one-layer \qaspa\ with depth-one \qaoa\ by exact state-vector
simulation.  The protocol uses seed \texttt{42}, evaluates connected
backbone-augmented random graphs with additional-edge probabilities
$p_{\mathrm{edge}}\in\{0.1,0.3,0.5,0.7\}$, and adds nineteen custom unweighted
topologies.  The \qaspa\ score is reported after the algorithm's classical
selection stage: the four leading QASPA bit strings are evaluated by the
Max-Cut objective, and the best cut is retained.  The QAOA score is the expected
approximation ratio after local optimization, reported as a best-of-$k$ curve
over one to ten restarts.  In these small exact simulations, fixed-angle
\qaspa\ produces reported group-mean selection ratios between $0.955$ and
$1.000$, the topology-level readout is unchanged in the retained finite-shot
checks under the rule $\lceil 0.03047\cdot 2^n+1016.2\rceil$, and depth-one
\qaoa\ improves with additional classical restarts while remaining lower than
the QASPA selection score in the reported mean summaries.  This comparison
motivates further study of the classical search budget needed to obtain
competitive Max-Cut samples at shallow circuit depth.
\end{abstract}

\section{Introduction}

The Quantum Approximate Optimization Algorithm (QAOA) introduced by Farhi,
Goldstone, and Gutmann is one of the canonical variational approaches to
combinatorial optimization on gate-model quantum computers
\cite{Farhi2014}.  For Max-Cut, the usual construction alternates a diagonal
problem unitary generated by the cut Hamiltonian and a transverse-field mixer.
The circuit has a direct graph interpretation, but it also exposes a practical
bottleneck: the reported performance depends not only on the quantum circuit
depth, but also on the classical search over variational angles.

In the noisy intermediate-scale quantum (NISQ) regime, this bottleneck becomes a
resource-accounting issue.  Increasing depth can improve the variational family
in principle, while gate noise, limited connectivity, and repeated circuit
evaluations limit what can be run in practice.  Recent reviews and performance
studies consequently report approximation ratio together with the number of
parameter evaluations, restarts, shots, and wall-clock time needed to obtain a
good solution \cite{Blekos2024,Lubinski2023,Dupont2024}.  The present study
isolates the restart effect by restricting the QASPA construction to unweighted
graphs.  Weighted Max-Cut would introduce a separate parameter-scaling question,
so it is treated here as related work rather than as the operating regime of the
proposed circuit
\cite{Shaydulin2022,Sureshbabu2024,Boy2024}.

The present work studies \qaspa, a Max-Cut circuit based on edge-local
controlled $R_y$ rotations.  A preliminary description of QASPA for Max-Cut
appears in \cite{Sapozhnyk2025}.  The present paper develops that construction
as a mathematical and experimental object suitable for comparison with
depth-one QAOA under a controlled amount of classical optimization.  The main
experimental variable is the number $k$ of independent QAOA optimizer restarts:
one QAOA run means one local optimization from one random initial point, and
the reported ``best of $k$'' value is the best expected approximation ratio
found after $k=1,\ldots,10$ such starts.

The paper makes three concrete additions.  First, it gives the exact algebra of the
$\mathrm{CX}$--$R_y$--$\mathrm{CX}$ block and states precisely how it acts on cut
and uncut edge configurations.  Second, it adds an unweighted four-vertex worked
example with a generated graph figure, edge-by-edge transformations, and output
probabilities.  Third, it provides a reproducible experimental comparison in the
form of explicit state-vector routines, table values, and code fragments that
connect the reported figures to exact simulations.

The main result is stated under these assumptions.  All graphs are unweighted,
all reported simulations are exact state-vector simulations on small $n=8$
instances except for the four-vertex worked example, and QASPA is run at the
fixed angle $\alpha=\pi/4$.  Within this scope, the observed fixed-angle QASPA
state places large-cut candidates inside the set chosen by its classical
selection stage in the tested random and custom graph groups, whereas depth-one
QAOA improves gradually as additional independent classical optimizer starts
are allowed.  The result is a controlled statement about
candidate quality and outer-loop optimization budget at shallow depth, not a
claim of asymptotic separation or hardware-level superiority.

\section{Related Work}

The Max-Cut problem asks for a bipartition of the vertices of an unweighted
graph that maximizes the number of crossing edges.  The Goemans--Williamson
semidefinite programming algorithm remains the standard approximation baseline,
with a worst-case expected approximation factor of approximately $0.87856$
\cite{Goemans1995}.  Modern exact and heuristic classical solvers are also
well developed, which is why quantum Max-Cut experiments need careful comparison
against classical baselines and against realistic execution costs
\cite{Rehfeldt2023,Lubinski2023,Dupont2024}.

Several QAOA variants aim to improve low-depth performance by changing the
mixer, the phase separator, or the ansatz.  Multi-angle QAOA adds
edge- and qubit-dependent angles and can match deeper standard QAOA on some
Max-Cut instances \cite{Herrman2022}.  Free-axis and expressive QAOA variants
allow the mixer to use $Y$ components rather than only $X$ rotations, improving
low-depth expressivity at the cost of additional parameters
\cite{Govia2021,Vijendran2024}.  Constraint-preserving and $XY$ mixers provide
another route for problems where the feasible subspace should be preserved by
construction \cite{Wang2020,Fuchs2022,Hadfield2023}.  These results motivate a
central design question for \qaspa: can some of the instance structure be moved
from a difficult outer-loop search into edge-local circuit structure?

Parameter setting is now a research topic in its own right.  Warm-started QAOA
uses classical relaxations or approximate solutions to bias the initial quantum
state and can improve low-depth behavior \cite{Egger2021,Tate2023,Yu2025}.
Other studies show that parameters can sometimes transfer across related
Max-Cut instances, reducing the number of expensive circuit evaluations
\cite{Shaydulin2022,Falla2024}.  These works support the experimental design
used here: instead of reporting only the best QAOA value after an unspecified
amount of tuning, we expose how solution quality changes as the number of
classical optimization starts increases.

\section{Mathematical Formulation}

Let $G=(V,E)$ be an undirected unweighted graph with $n=|V|$.  A bit string
$z\in\{0,1\}^n$ defines a cut, and the cut value is
\begin{equation}
  C_G(z)
  =
  \sum_{(i,j)\in E} (z_i\oplus z_j).
  \label{eq:cut-binary}
\end{equation}
Equivalently, with spin variables $s_i=(-1)^{z_i}$,
\begin{equation}
  C_G(z)
  =
  \frac{1}{2}\sum_{(i,j)\in E}(1-s_i s_j).
  \label{eq:cut-spin}
\end{equation}
The usual Max-Cut cost Hamiltonian is therefore
\begin{equation}
  H_C
  =
  \frac{1}{2}\sum_{(i,j)\in E}(I-Z_iZ_j),
  \label{eq:cost-hamiltonian}
\end{equation}
whose eigenvalue on computational basis state $|z\rangle$ is $C_G(z)$.

Depth-one QAOA prepares
\begin{equation}
  |\psi_{\mathrm{QAOA}}(\gamma,\beta)\rangle
  =
  U_B(\beta)U_C(\gamma)\ketplus^{\otimes n},
  \label{eq:qaoa-state}
\end{equation}
where
\begin{equation}
  \begin{aligned}
  U_C(\gamma)&=e^{-i\gamma H_C},\\
  U_B(\beta)&=e^{-i\beta\sum_i X_i}\\
  &=\prod_{i=0}^{n-1}R_x^{(i)}(2\beta).
  \end{aligned}
  \label{eq:qaoa-unitaries}
\end{equation}
The classical optimizer searches over $(\gamma,\beta)$ to maximize
$\langle H_C\rangle$.

\qaspa\ keeps the initial Hadamard layer but replaces the global $X$ mixer with
edge-local $Y$-axis rotations.  For each unweighted edge $(i,j)$ we set
\begin{equation}
  \theta_{ij}=\alpha,
  \label{eq:edge-angle}
\end{equation}
where $\alpha$ is a global edge angle.  No edge weights are introduced in the
current QASPA circuit, so the graph contributes through its unweighted edge set
and through the order in which the corresponding blocks are applied.  The QASPA
edge block is
\begin{equation}
  U_{ij}^{(Y)}(\theta_{ij})
  =
  \mathrm{CX}_{i,j}\,
  R_y^{(j)}(\theta_{ij})\,
  \mathrm{CX}_{i,j}.
  \label{eq:qaspa-edge-block}
\end{equation}
For one layer and a fixed edge ordering, the circuit is
\begin{equation}
  |\psi_{\mathrm{QASPA}}(\alpha)\rangle
  =
  \prod_{(i,j)\in E}^{\rightarrow}
  U_{ij}^{(Y)}(\alpha)\,
  \ketplus^{\otimes n}.
  \label{eq:qaspa-state}
\end{equation}
The arrow reminds us that these edge blocks do not generally commute when
edges share a vertex.

The algebra of the edge block is explicit because
$R_y(\theta)=\exp(-i\theta Y/2)$ and $X R_y(\theta)X=R_y(-\theta)$ imply
\begin{align}
  U_{ij}^{(Y)}(\theta)
  &=
  |0\rangle\langle 0|_i\otimes R_y^{(j)}(\theta)
  +
  |1\rangle\langle 1|_i\otimes R_y^{(j)}(-\theta) \notag\\
  &=
  \exp\!\left(-\frac{i\theta}{2}Z_iY_j\right).
  \label{eq:edge-exponential}
\end{align}
Thus the control qubit selects the sign of the $Y$ rotation on the target.
Writing $c=\cos(\theta/2)$ and $s=\sin(\theta/2)$, the computational-basis
action is
\begin{align}
  |00\rangle &\mapsto c|00\rangle+s|01\rangle,&
  |01\rangle &\mapsto -s|00\rangle+c|01\rangle, \notag\\
  |10\rangle &\mapsto c|10\rangle-s|11\rangle,&
  |11\rangle &\mapsto s|10\rangle+c|11\rangle.
  \label{eq:edge-basis-action}
\end{align}
Each control-conditioned two-dimensional subspace contains one cut and one
uncut configuration for the edge.  The block therefore redistributes amplitude
between cut and uncut states, rather than only applying a diagonal phase.  Up
to a target-qubit basis rotation that maps $Y_j$ to $Z_j$, the generator
$Z_iY_j$ is equivalent to a $Z_iZ_j$ edge interaction.

\begin{figure}[t]
  \centering
  \includegraphics[width=\linewidth]{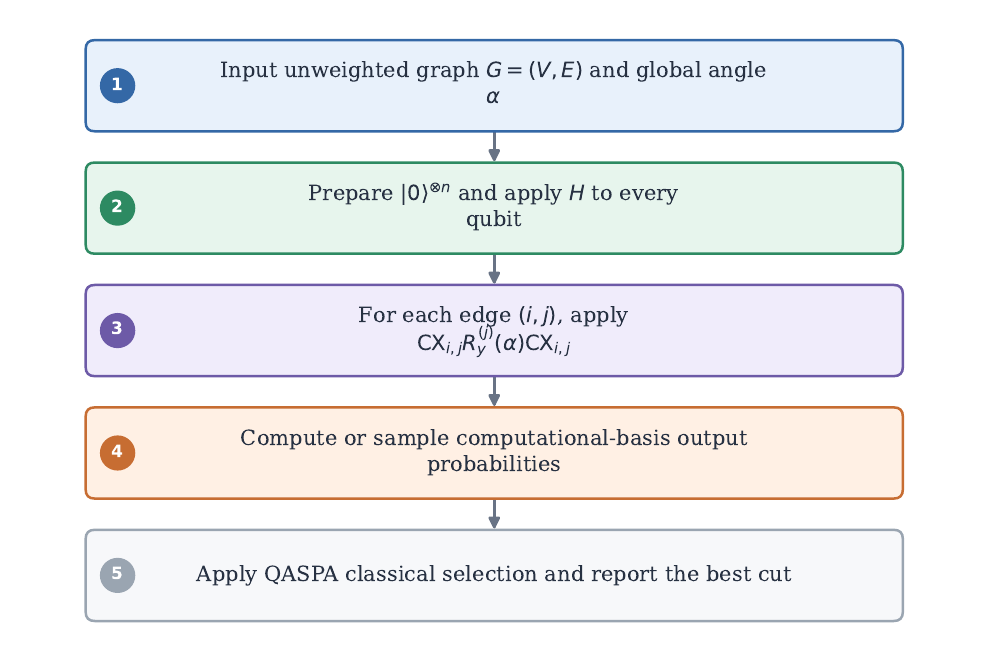}
  \caption{Block scheme of the one-layer QASPA-MaxCut procedure for
  unweighted graphs.}
  \label{fig:qaspa-flow}
\end{figure}

To connect the algebraic block with the actual computation,
Figure~\ref{fig:qaspa-flow} presents the one-layer QASPA-MaxCut construction as
a pipeline.  The graph first determines the edge-block schedule, the scheduled
blocks define the state-vector evolution, and the resulting probabilities are
then passed to the classical selection stage.

Listing~\ref{lst:qaspa-core} uses the same order in code form, so the transition
from the diagram to the implementation preserves the notation used in
Eqs.~\eqref{eq:qaspa-edge-block}--\eqref{eq:edge-basis-action}.

\begin{lstlisting}[style=paperpseudo,caption={Implementation skeleton for the QASPA edge layer.},label={lst:qaspa-core}]
state = zero_state(2**n)
state[0] = 1
for q in vertices:
    apply_h(state, q)

for (i, j) in edges:
    apply_cx(state, i, j)
    apply_ry(state, j, alpha)
    apply_cx(state, i, j)

probabilities = abs(state)**2
expected_cut = sum(probabilities[z] * cut_value(z, edges) for z in states)
\end{lstlisting}

\section{Circuit Construction Details}

The mathematical definition above is compact, but an implementation needs a
fixed interpretation of qubit order, edge order, and output bit strings.  This
paper uses the q0-first convention throughout: the leftmost character of a
reported bit string is the value of qubit $0$.  The convention is not a
mathematical restriction, but stating it avoids a common source of mistakes
when comparing simulator output, tables, and graph partitions.

\begin{figure}[t]
  \centering
  \includegraphics[width=\linewidth]{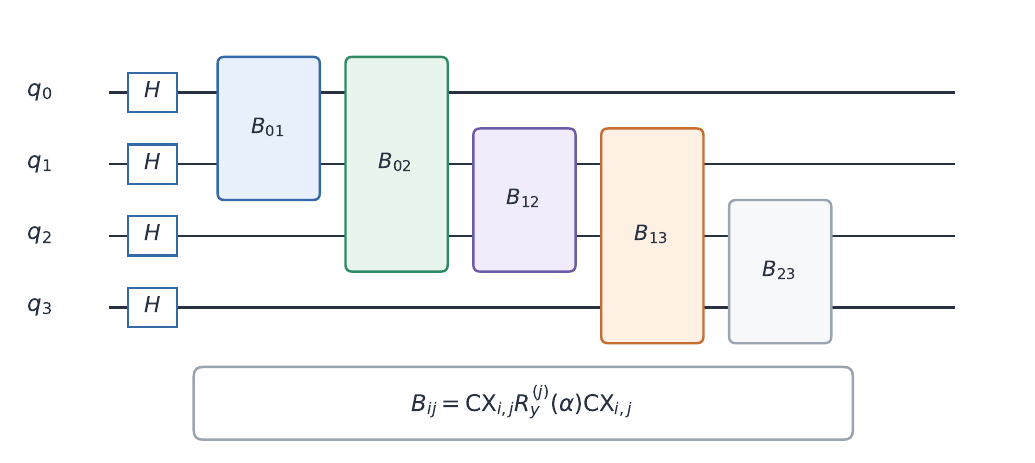}
  \caption{Circuit schedule for the four-vertex example, where each edge block
  is a controlled sign-selecting $R_y$ operation.}
  \label{fig:circuit-schedule}
\end{figure}

Figure~\ref{fig:circuit-schedule} expands the single product in
Equation~\eqref{eq:qaspa-state} into an algorithmic circuit schedule rather
than a hardware transpilation.  Each graph edge is translated into one local
block, and the displayed order is part of the experimental definition: if two
adjacent edge blocks share a qubit, changing their order can change the final
state.  The experiment therefore fixes a deterministic edge order before any
probability or expectation value is computed.

\begin{figure}[t]
  \centering
  \includegraphics[width=\linewidth]{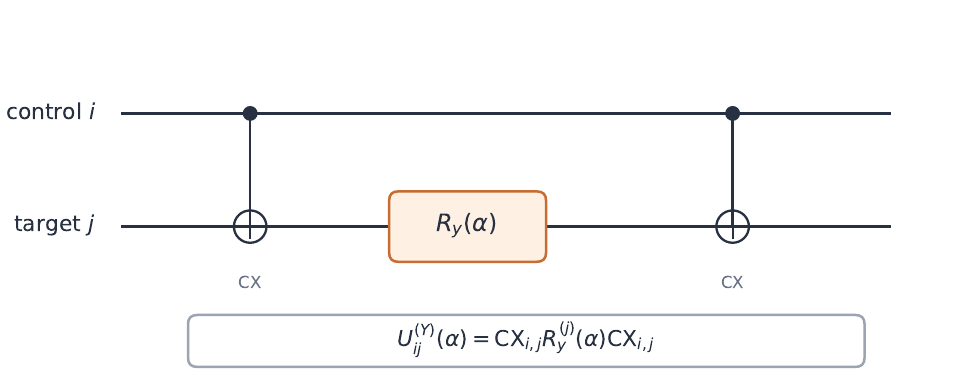}
  \caption{Gate-level structure of one QASPA edge block used to encode a graph
  edge in the circuit.}
  \label{fig:edge-block-detail}
\end{figure}

The edge block itself is shown in Figure~\ref{fig:edge-block-detail}.  The
first CNOT maps the parity information of the pair into the target qubit, the
$R_y$ gate rotates that target, and the final CNOT restores the computational
labeling.  This is the operational reason why the block is more than a
diagonal phase: it modifies amplitudes in a two-dimensional subspace attached
to an edge.

\begin{figure}[t]
  \centering
  \includegraphics[width=\linewidth]{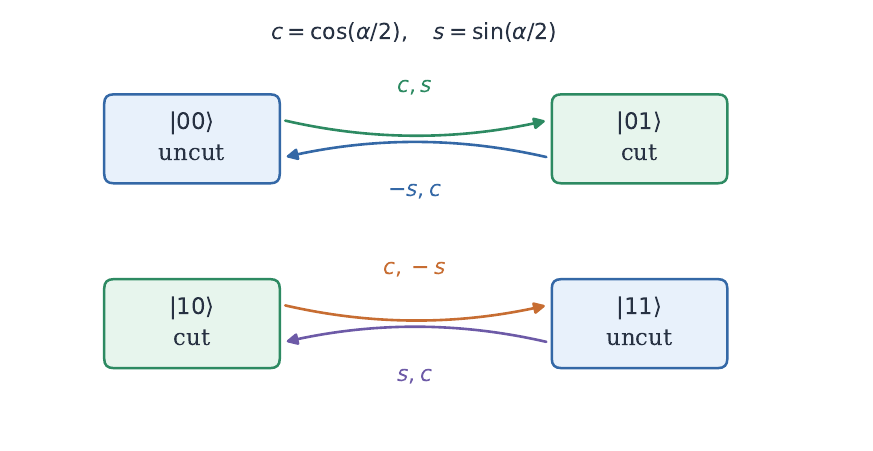}
  \caption{Amplitude redistribution induced by one edge block, with
  $c=\cos(\alpha/2)$ and $s=\sin(\alpha/2)$ defining the local rotation
  coefficients.}
  \label{fig:amplitude-pairs}
\end{figure}

Figure~\ref{fig:amplitude-pairs} gives the same operation as an amplitude-flow
picture.  For a fixed value of the control qubit, the target qubit is rotated
inside a pair of basis states, where one state cuts the edge and the other does
not.  The block therefore changes how much probability is assigned to cut and
uncut configurations before the final measurement.

The bit-string convention introduced above becomes concrete in
Listing~\ref{lst:cut-value}.  The same cut-value routine is used for the test
graph, the exact optimum, and the expectation ratio, which keeps the
mathematical definition and the numerical reporting aligned.

\begin{lstlisting}[style=paperpseudo,caption={Cut-value routine for q0-first bit strings.},label={lst:cut-value}]
def bit_at(mask, q):
    return (mask >> q) & 1

def cut_value(mask, edges):
    total = 0
    for i, j in edges:
        if bit_at(mask, i) != bit_at(mask, j):
            total += 1
    return total

def bitstring(mask, n):
    return "".join(str(bit_at(mask, q)) for q in range(n))
\end{lstlisting}

With the cut convention fixed, the expectation value can be evaluated by direct
enumeration of the computational basis.  This exact state-vector calculation is
practical for the small graphs used in this article, and it removes
sampling noise from the methodological comparison.  Listing~\ref{lst:expectation}
implements the same quantity later normalized in
Equation~\eqref{eq:expected-ratio}.

\begin{lstlisting}[style=paperpseudo,caption={Exact expected cut from a state vector.},label={lst:expectation}]
def expected_cut(state, edges):
    probabilities = abs(state)**2
    value = 0.0
    for mask, probability in enumerate(probabilities):
        value += probability * cut_value(mask, edges)
    return value
\end{lstlisting}

\section{Worked Test Graph}

\begin{figure}[t]
  \centering
  \includegraphics[width=0.58\linewidth]{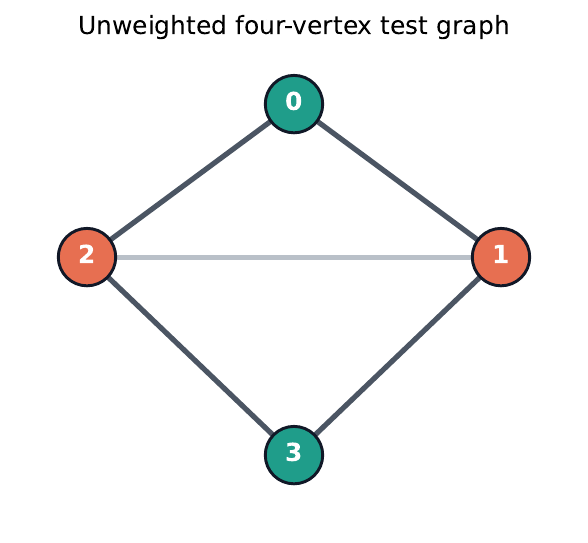}
  \caption{Unweighted test graph, where the dark edges cross the optimal
  partition \texttt{0110} and the light edge does not.}
  \label{fig:test-graph}
\end{figure}

Figure~\ref{fig:test-graph} shows the unweighted four-vertex graph used as a
mathematical example.  The optimum cut value is $C_G^\ast=4$, obtained by the
q0-first bit strings \texttt{0110} and \texttt{1001}.  The colors in the figure
show one of these optimal partitions.

For this graph, the initial state and the first QASPA layer become explicit:
\begin{align}
  |\psi_0\rangle
    &= |0000\rangle, \notag\\
  |\psi_H\rangle
    &= H^{\otimes 4}|\psi_0\rangle
     = \frac{1}{4}\sum_{z\in\{0,1\}^4}|z\rangle, \label{eq:test-hadamard}\\
  |\psi_{\mathrm{QASPA}}\rangle
    &=
    U_{23}^{(Y)}(\alpha)
    U_{13}^{(Y)}(\alpha)
    U_{12}^{(Y)}(\alpha) \notag\\
    &\quad\times
    U_{02}^{(Y)}(\alpha)
    U_{01}^{(Y)}(\alpha)
    |\psi_H\rangle . \label{eq:test-layer}
\end{align}
The fixed-angle condition $\alpha=\pi/4$ assigns the same rotation scale to each
unweighted edge, as recorded in Table~\ref{tab:test-edge-angles}.

\begin{table}[b]
  \centering
  \caption{Edge-specific QASPA angles for the test graph at $\alpha=\pi/4$.}
  \label{tab:test-edge-angles}
  \begin{tabular}{lcc}
    \toprule
    Edge & Symbolic angle $\theta_{ij}$ & Numeric angle \\
    \midrule
    $(0,1)$ & $\alpha$ & 0.7854 \\
    $(0,2)$ & $\alpha$ & 0.7854 \\
    $(1,2)$ & $\alpha$ & 0.7854 \\
    $(1,3)$ & $\alpha$ & 0.7854 \\
    $(2,3)$ & $\alpha$ & 0.7854 \\
    \bottomrule
  \end{tabular}
\end{table}

At this fixed global scale, the five edge transformations are
\begin{align}
  U_{01}^{(Y)}(\alpha)
    &=
    \exp\!\left(-\frac{i\alpha}{2}Z_0Y_1\right), \notag\\
  U_{02}^{(Y)}(\alpha)
    &=
    \exp\!\left(-\frac{i\alpha}{2}Z_0Y_2\right), \notag\\
  U_{12}^{(Y)}(\alpha)
    &=
    \exp\!\left(-\frac{i\alpha}{2}Z_1Y_2\right), \notag\\
  U_{13}^{(Y)}(\alpha)
    &=
    \exp\!\left(-\frac{i\alpha}{2}Z_1Y_3\right), \notag\\
  U_{23}^{(Y)}(\alpha)
    &=
    \exp\!\left(-\frac{i\alpha}{2}Z_2Y_3\right),
    \qquad
  \alpha=\frac{\pi}{4}.
  \label{eq:test-edge-transformations}
\end{align}
Equations~\eqref{eq:test-layer} and~\eqref{eq:test-edge-transformations} give
the complete symbolic circuit for the example.

Table~\ref{tab:test-top-states} shows the largest output probabilities after
the full QASPA layer at $\alpha=\pi/4$.  For this unweighted example, the two
most probable strings are exactly the two optimum cuts.  This provides a
limited sanity check for the construction: in this instance, the edge-encoded
$R_y$ layer moves probability mass toward large cut values, even though a fixed
angle is not a general guarantee of exact optimization.

\begin{table*}[t]
  \centering
  \caption{Top QASPA output states for the test graph at $\alpha=\pi/4$, with
  q0-first bit strings matching the simulator convention.}
  \label{tab:test-top-states}
  \begin{tabular}{ccccc}
    \toprule
    Rank & Bit string & Probability & Cut value & Optimal? \\
    \midrule
    1 & \texttt{0110} & 0.2134 & 4.00 & yes \\
    2 & \texttt{1001} & 0.2134 & 4.00 & yes \\
    3 & \texttt{0100} & 0.1067 & 3.00 & no \\
    4 & \texttt{1010} & 0.1067 & 3.00 & no \\
    5 & \texttt{0101} & 0.1067 & 3.00 & no \\
    6 & \texttt{1011} & 0.1067 & 3.00 & no \\
    7 & \texttt{1100} & 0.0366 & 3.00 & no \\
    8 & \texttt{0010} & 0.0366 & 3.00 & no \\
    \bottomrule
  \end{tabular}
\end{table*}

\begin{figure}[t]
  \centering
  \includegraphics[width=\linewidth]{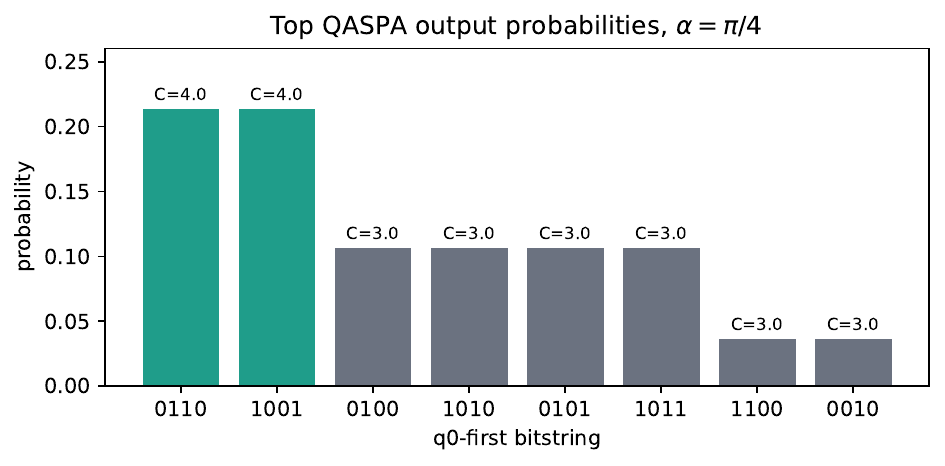}
  \caption{Largest output probabilities for the unweighted test graph under one
  QASPA layer with $\alpha=\pi/4$, where green bars mark optimal cuts.}
  \label{fig:test-distribution}
\end{figure}

The probability chart in Figure~\ref{fig:test-distribution} complements the
exact values in Table~\ref{tab:test-top-states}.  The table identifies the two
optimal strings as the most probable outputs, while the figure shows that the
fixed-angle layer still leaves visible probability mass on nearby nonoptimal
strings.  This observation sets up the benchmark interpretation: QASPA is
treated as a candidate generator, and the reported score asks whether the
retained high-probability candidates contain a large cut.

Because the four-vertex graph is small, the product in
Equation~\eqref{eq:test-layer} can be unfolded into an edge-by-edge trace.
Table~\ref{tab:edge-trace} gives the simulator order explicitly, connecting the
graph, the circuit schedule, and the implementation convention before the paper
moves to the numerical protocol.  This deterministic trace is part of the
algorithmic specification, since edge blocks acting on overlapping qubits are
not treated as interchangeable algebraic factors in the reported experiments.

\begin{center}
  \small
  \captionof{table}{Trace of the QASPA edge schedule for the four-vertex graph.}
  \label{tab:edge-trace}
  \begin{tabular}{clll}
    \toprule
    Step & Edge & Operator & Purpose \\
    \midrule
    1 & $(0,1)$ & $U_{01}^{(Y)}(\alpha)$ & First relation \\
    2 & $(0,2)$ & $U_{02}^{(Y)}(\alpha)$ & Second relation at $0$ \\
    3 & $(1,2)$ & $U_{12}^{(Y)}(\alpha)$ & Middle relation \\
    4 & $(1,3)$ & $U_{13}^{(Y)}(\alpha)$ & Lower-right relation \\
    5 & $(2,3)$ & $U_{23}^{(Y)}(\alpha)$ & Lower-left relation \\
    \bottomrule
  \end{tabular}
\end{center}

Figure~\ref{fig:edge-interpretation-chain} extends the edge-level trace to the
final QASPA selection score, and Listing~\ref{lst:test-graph-manual} ends the
worked example by writing the four-vertex graph in the same form used by the
later experiments.

\noindent\begin{minipage}{\linewidth}
  \centering
  \includegraphics[width=\linewidth]{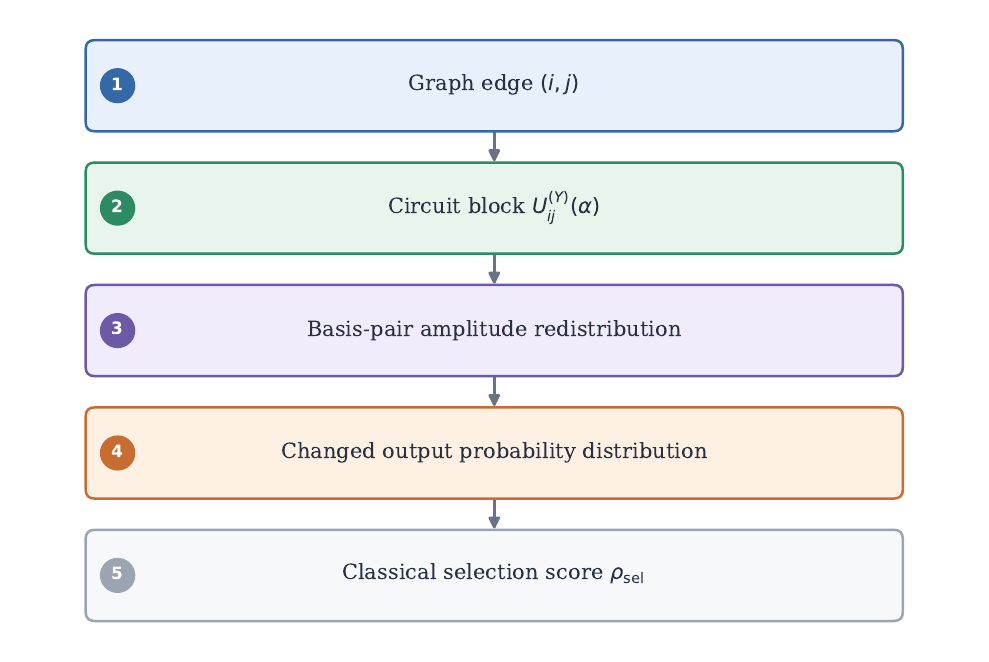}
  \captionof{figure}{Interpretation chain from a graph edge to its contribution in the
  QASPA classical selection score.}
  \label{fig:edge-interpretation-chain}
\end{minipage}

\noindent\begin{minipage}{\linewidth}
\begin{lstlisting}[style=paperpseudo,caption={Manual construction of the four-vertex example.},label={lst:test-graph-manual}]
edges = [
    (0, 1),
    (0, 2),
    (1, 2),
    (1, 3),
    (2, 3),
]

alpha = pi / 4
state = qaspa_state(edges, n=4, alpha=alpha)
top_states = sort_by_probability(state)[:8]
optimum, optimum_masks = exact_optimum(edges, n=4)
\end{lstlisting}
\end{minipage}

Together, Figure~\ref{fig:circuit-schedule}, Table~\ref{tab:edge-trace},
Figure~\ref{fig:edge-interpretation-chain}, and
Listing~\ref{lst:test-graph-manual} connect the visual circuit placement with
the algebraic reason for each block and the final computational routine.  The
QASPA circuit first redistributes amplitudes through the ordered edge blocks,
the resulting probability distribution is sampled or sorted, and only then are
the four leading bit strings evaluated by the Max-Cut objective.  This
distinction matters because the QASPA value reported in the tables is the score
after the algorithm's classical selection stage, while the QAOA value is an
expected approximation ratio optimized over restart budgets.

As a result, the example is not a hand-calculated special case detached from the
experimental protocol.  The same routines that evaluate the random graphs also
evaluate the small graph in Figure~\ref{fig:test-graph}, allowing the paper to
move from a transparent trace to the larger benchmark without changing the
computational definitions.

\section{Experimental Protocol and Results}

The worked example fixes the conventions; the experiment now asks how
fixed-angle \qaspa\ compares with depth-one QAOA when the classical optimization
budget is made visible.  The QASPA condition uses $\alpha=\pi/4$ without an
additional outer-loop angle search, while the QAOA condition uses a local
Nelder--Mead optimizer.  In the QAOA condition, one optimization run means one
random initialization of $(\gamma,\beta)$ followed by local optimization, and
for each graph we record the best expected approximation ratio after
$k=1,\ldots,10$ independent starts.

All simulations use exact state vectors and exact Max-Cut optima, and every
graph in the experiment is unweighted.  The random graphs are connected
backbone-augmented instances: a path backbone is inserted first, and additional
nonbackbone edges are then sampled independently.  Each additional-edge
probability group uses $n=8$
vertices and $12$ instances, with additional-edge probabilities
$p_{\mathrm{edge}}\in\{0.1,0.3,0.5,0.7\}$.  A separate custom group contains
nineteen hand-specified connected topologies on the same number of vertices.  The
local QAOA optimizer uses the exact expected cut
\begin{equation}
  \rho_{\mathrm{exp}}
  =
  \frac{\mathbb{E}_{z\sim |\psi|^2}[C_G(z)]}{C_G^\ast}.
  \label{eq:expected-ratio}
\end{equation}
The reported QASPA score is the output of the method's classical selection
stage.  Let $S_{\mathrm{QASPA}}(\psi)$ be the four computational-basis bit
strings selected from the largest output probabilities under the fixed-angle
QASPA state $\psi$.  The QASPA selection ratio is
\begin{equation}
  \rho_{\mathrm{sel}}(\psi)
  =
  \frac{\max_{z\in S_{\mathrm{QASPA}}(\psi)} C_G(z)}{C_G^\ast}.
  \label{eq:qaspa-selection-ratio}
\end{equation}
QASPA is therefore evaluated as a candidate-generating algorithm whose
classical selection stage keeps the best cut among the selected candidates.
QAOA is evaluated by the expected approximation ratio in
Equation~\eqref{eq:expected-ratio}, because the comparison is designed to show
how much the variational baseline improves with repeated classical
optimization.  The result section keeps this distinction explicit: QASPA asks
whether a small retained candidate set already contains a large cut, while QAOA
asks how the optimized expectation improves as more classical starts are spent
on the same graph group.  To make the metric difference auditable,
Table~\ref{tab:metric-control} also reports the fixed-angle QASPA expected
ratio alongside the prescribed QASPA selection score.

To connect this exact readout rule with a finite-sampling setting, the
topology-specific QASPA test also uses the shot budget
\begin{equation}
  N_{\mathrm{shots}}(n)
  =
  \left\lceil 0.03047\cdot 2^n + 1016.2 \right\rceil .
  \label{eq:shot-rule}
\end{equation}
For the $n=8$ graphs in this manuscript, Equation~\eqref{eq:shot-rule} gives
$N_{\mathrm{shots}}=1025$.  The sampled readout test draws that many outcomes
from the QASPA output distribution, selects the four most frequently observed
bit strings, and reports the best Max-Cut value after this classical selection
stage.
This readout is repeated $50$ times for each custom topology with a
deterministic sampling stream derived from seed \texttt{42}.  The shot budget is
therefore adaptive in the number of qubits rather than a single constant budget
reused across all graph sizes, and the reported sampled values include the
variability of the finite readout.  Listing~\ref{lst:shot-count-rule} records
the rule in the same pseudocode style as the algorithmic fragments below.

\begin{lstlisting}[style=paperpseudo,caption={Adaptive QASPA shot-count rule used for finite readout.},label={lst:shot-count-rule}]
def qaspa_shot_count(n):
    return ceil(0.03047 * 2**n + 1016.2)
\end{lstlisting}

Table~\ref{tab:protocol-controls} collects the fixed choices behind the
experiment and makes the restart budget visible.  The comparison is between a
fixed QASPA edge angle and a QAOA optimizer that receives a controlled number
of independent starts, rather than between arbitrary tuning effort and a fixed
circuit.

\begin{table*}[t]
  \centering
  \footnotesize
  \caption{Protocol controls used to make the numerical comparison reproducible
  and to expose the classical restart budget.}
  \label{tab:protocol-controls}
  \begin{tabular}{ll}
    \toprule
    Item & Value used in the manuscript \\
    \midrule
    Graph type & Connected unweighted random backbones; custom topologies \\
    Graph size & $n=8$ vertices \\
    Additional-edge probabilities & $p_{\mathrm{edge}}\in\{0.1,0.3,0.5,0.7\}$ \\
    Instances per additional-edge probability & $12$ \\
    Custom topology count & $19$ \\
    QASPA fixed angle & $\alpha=\pi/4$ \\
    QAOA depth & One alternating layer \\
    QAOA restart budgets & $k=1,\ldots,10$ independent starts \\
    Optimizer & Nelder--Mead, maximum iteration count $80$ \\
    Reported scores & QASPA selection ratio; QAOA best expected ratio \\
    QASPA shot budget & $N_{\mathrm{shots}}=\left\lceil 0.03047\cdot 2^n+1016.2\right\rceil=1025$ for $n=8$ \\
    Random seed & \texttt{42} \\
    \bottomrule
  \end{tabular}
\end{table*}

Table~\ref{tab:qaoa-optimizer-diagnostics} reports the behavior of the local
QAOA optimizer under these controls.  The returned state from every local start
is evaluated, including starts where the optimizer terminates at the iteration
limit rather than satisfying its convergence criterion.  This policy avoids
silently discarding poor local starts, while the success column records how
often the local search terminates normally.

\begin{table*}[t]
  \centering
  \caption{Diagnostics for the depth-one QAOA local optimizer.  Each local start
  uses Nelder--Mead with maximum iteration count $80$; $n_{\mathrm{fev}}$ denotes
  objective-function evaluations.}
  \label{tab:qaoa-optimizer-diagnostics}
  \small
  \begin{tabular}{lccccc}
    \toprule
    Graph group & Starts & Successful & Success rate & Mean $n_{\mathrm{fev}}$ & Max $n_{\mathrm{fev}}$ \\
    \midrule
    Backbone $p=0.1$ & 120 & 120 & 1.000 & 115.1 & 155 \\
Backbone $p=0.3$ & 120 & 119 & 0.992 & 112.3 & 154 \\
Backbone $p=0.5$ & 120 & 119 & 0.992 & 114.0 & 152 \\
Backbone $p=0.7$ & 120 & 117 & 0.975 & 117.8 & 159 \\
Custom topologies & 190 & 187 & 0.984 & 115.2 & 153 \\
\bottomrule

  \end{tabular}
\end{table*}

After these controls are fixed, Figure~\ref{fig:benchmark-flow} summarizes the
same protocol as a block scheme.  The diagram prepares the interpretation of
Figure~\ref{fig:restart-profile}: the QAOA curve is a best-so-far curve over
the restart budget, while the QASPA reference line is computed once per graph
under the fixed-angle condition.

\noindent\begin{minipage}{\linewidth}
  \centering
  \includegraphics[width=\linewidth]{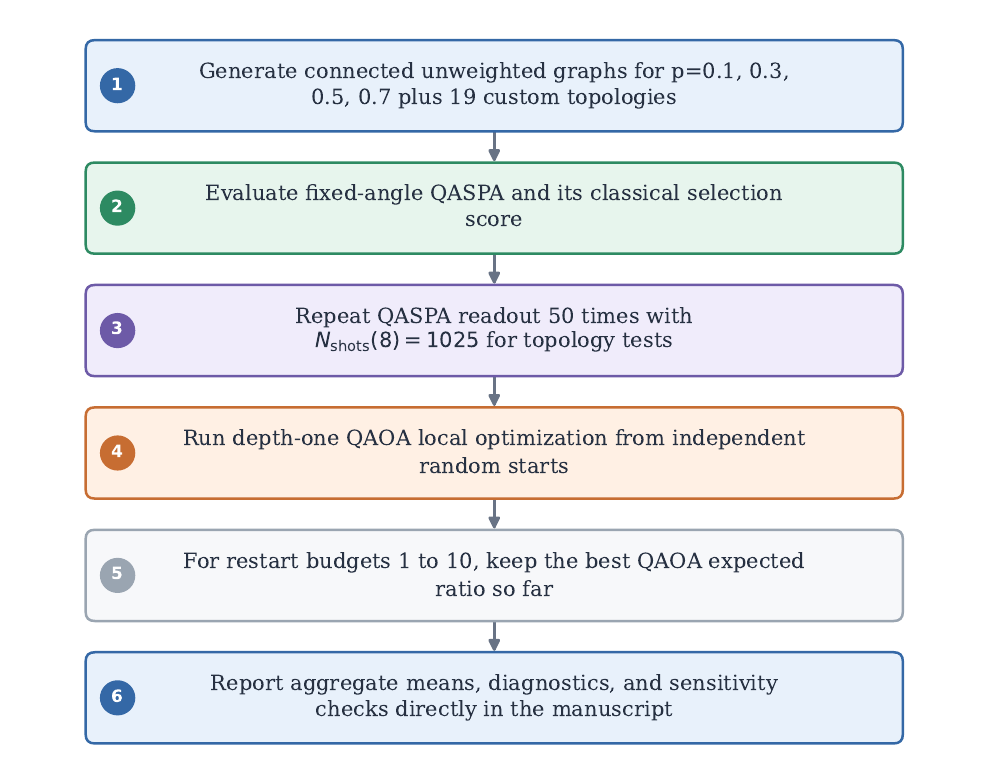}
  \captionof{figure}{Block scheme of the experimental protocol, with numerical aggregates
  reported directly in the manuscript tables.}
  \label{fig:benchmark-flow}
\end{minipage}

At the implementation level, Listing~\ref{lst:restart-core} gives the compact
wrapper behind the restart curve.  QASPA is evaluated once at the fixed angle,
whereas QAOA is updated as a best-so-far value over repeated local optimizer
starts.

\begin{lstlisting}[style=paperpseudo,float,caption={Implementation skeleton for the restart comparison.},label={lst:restart-core}]
def qaspa_selection_ratio(state, graph):
    probabilities = abs(state)**2
    selected = argsort_descending(probabilities)[:4]
    best_cut = max(cut_value(mask, graph.edges) for mask in selected)
    return best_cut / exact_optimum(graph)

for graph in random_graphs + custom_graphs:
    qaspa_fixed = qaspa_selection_ratio(qaspa_state(graph, alpha=pi/4), graph)

    best_qaoa = -infinity
    for k in range(1, 11):
        gamma, beta = random_initial_angles()
        state = local_optimize_qaoa(graph, gamma, beta)
        value = expected_ratio(state, graph)
        best_qaoa = max(best_qaoa, value)
        report_best_of_k(graph, k, best_qaoa, qaspa_fixed)
\end{lstlisting}

The restart profile in Figure~\ref{fig:restart-profile} is the main visual
summary of this protocol.  It uses seed \texttt{42}, and the aggregate values
needed to read the curves are printed directly in the manuscript tables.

\begin{center}
  \begin{minipage}{\linewidth}
    \centering
    \includegraphics[width=\linewidth]{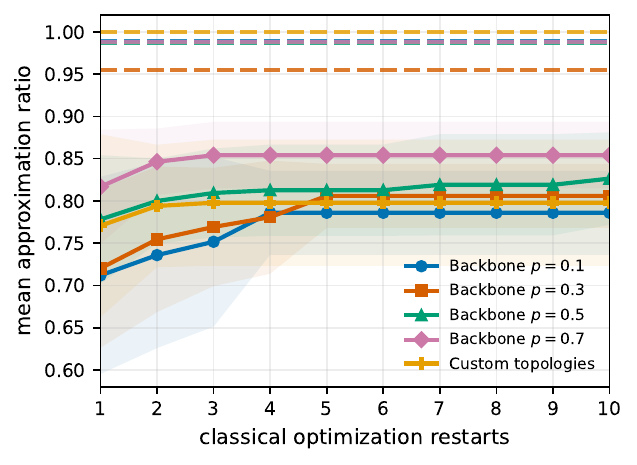}
    \captionof{figure}{Mean approximation ratio for depth-one QAOA as a function
    of the number of independent classical optimizer restarts.  Colors distinguish
    graph groups; solid curves show the QAOA best-of-$k$ expected ratio, and
    dashed horizontal lines show the corresponding fixed-angle QASPA classical
    selection score.  Light shaded regions show one sample standard deviation
    over the graph instances in each group.}
    \label{fig:restart-profile}
  \end{minipage}
\end{center}

Tables~\ref{tab:metric-control}--\ref{tab:benchmark-summary-custom} give the
numerical values behind Figure~\ref{fig:restart-profile}.  The expanded
experiment shows a similar restart-budget pattern across low-, medium-, and
high-additional-edge-probability groups and across custom topologies.  For
example, in the reported means QAOA improves from $0.712$ to
$0.786$ on the $p_{\mathrm{edge}}=0.1$ group and from $0.771$ to $0.798$ on the
custom topology group, while the corresponding fixed QASPA selection values are
$0.990$ and $1.000$.

Table~\ref{tab:metric-control} adds a metric-control check at the largest QAOA
restart budget.  The QASPA selection column is
$\rho_{\mathrm{sel}}$, while the QASPA expected column applies
Equation~\eqref{eq:expected-ratio} to the full fixed-angle QASPA distribution.
This separates the comparison based on the prescribed QASPA readout from the
comparison based on an expectation-value metric.  The control reduces the risk
that the QASPA selection scores are read as an expectation-value claim: in this
small experiment the fixed-angle QASPA expectation is above the ten-restart QAOA
expectation in the reported group means, but the margins are narrower than in
the QASPA selection column.  The later tables therefore keep the prescribed
QASPA selection score as the primary value while leaving the expectation
comparison visible as a calibration point.

\begin{table*}[t]
  \centering
  \caption{Metric-control comparison at the ten-restart QAOA budget.  The QASPA
  selection column is $\rho_{\mathrm{sel}}$ from
  Equation~\eqref{eq:qaspa-selection-ratio}; the two expected-ratio columns use
  Equation~\eqref{eq:expected-ratio}.}
  \label{tab:metric-control}
  \small
  \begin{tabular}{lccc}
    \toprule
    Graph group & QASPA selection $\rho_{\mathrm{sel}}$ & QASPA expected $\rho_{\mathrm{exp}}$ & QAOA best-of-10 \\
    \midrule
    Backbone $p=0.1$ & 0.990 & 0.858 & 0.786 \\
Backbone $p=0.3$ & 0.955 & 0.852 & 0.806 \\
Backbone $p=0.5$ & 0.987 & 0.858 & 0.826 \\
Backbone $p=0.7$ & 0.988 & 0.880 & 0.854 \\
Custom topologies & 1.000 & 0.884 & 0.798 \\
\bottomrule

  \end{tabular}
\end{table*}

\begin{table*}[t]
  \centering
  \caption{Experimental summary for sparse backbone-augmented graph groups.  QAOA
  values are means over graph instances of the best expected ratio after $k$
  optimizer restarts; QASPA values use the classical selection rule.}
  \label{tab:benchmark-summary}
  \scriptsize
  \begin{tabular}{ccccc}
    \toprule
    Graph group & Restarts $k$ & QAOA mean & QAOA std. & QASPA fixed \\
    \midrule
    Backbone $p=0.1$ & 1 & 0.712 & 0.117 & 0.990 \\
Backbone $p=0.1$ & 2 & 0.736 & 0.110 & 0.990 \\
Backbone $p=0.1$ & 3 & 0.752 & 0.100 & 0.990 \\
Backbone $p=0.1$ & 4 & 0.786 & 0.050 & 0.990 \\
Backbone $p=0.1$ & 5 & 0.786 & 0.050 & 0.990 \\
Backbone $p=0.1$ & 6 & 0.786 & 0.050 & 0.990 \\
Backbone $p=0.1$ & 7 & 0.786 & 0.050 & 0.990 \\
Backbone $p=0.1$ & 8 & 0.786 & 0.050 & 0.990 \\
Backbone $p=0.1$ & 9 & 0.786 & 0.050 & 0.990 \\
Backbone $p=0.1$ & 10 & 0.786 & 0.050 & 0.990 \\
Backbone $p=0.3$ & 1 & 0.720 & 0.094 & 0.955 \\
Backbone $p=0.3$ & 2 & 0.754 & 0.086 & 0.955 \\
Backbone $p=0.3$ & 3 & 0.769 & 0.070 & 0.955 \\
Backbone $p=0.3$ & 4 & 0.781 & 0.067 & 0.955 \\
Backbone $p=0.3$ & 5 & 0.806 & 0.038 & 0.955 \\
Backbone $p=0.3$ & 6 & 0.806 & 0.038 & 0.955 \\
Backbone $p=0.3$ & 7 & 0.806 & 0.038 & 0.955 \\
Backbone $p=0.3$ & 8 & 0.806 & 0.038 & 0.955 \\
Backbone $p=0.3$ & 9 & 0.806 & 0.038 & 0.955 \\
Backbone $p=0.3$ & 10 & 0.806 & 0.038 & 0.955 \\
\bottomrule

  \end{tabular}
\end{table*}

The sparse graph groups in Table~\ref{tab:benchmark-summary} provide the most
direct view of restart sensitivity when few nonbackbone edges are present.  The
largest gain comes from the first few QAOA restarts, after which the expected
ratio changes little and is lower than the fixed QASPA selection reference in
both low-probability group means.
Table~\ref{tab:benchmark-summary-dense} extends the same reading to the
medium- and high-probability random groups, where more nonbackbone edges are
available but the restart budget is kept unchanged.
Reading the sparse and denser random groups together separates two effects that
can otherwise be confused in a small Max-Cut study.  A larger value of
$p_{\mathrm{edge}}$ changes the number of cost terms, whereas the restart count
changes only the amount of classical search given to QAOA after the depth-one
circuit has been fixed.  The resulting plateau is therefore a restart-budget
diagnostic: it indicates that the chosen local optimizer and initialization
scheme stopped improving within ten starts, not that the instance family or
QAOA ansatz has been exhausted.  This keeps the interpretation local before the
analysis moves from randomly augmented backbones to hand-specified topologies.

\begin{table*}[t]
  \centering
  \caption{Experimental summary for medium- and high-probability backbone-augmented graph
  groups.}
  \label{tab:benchmark-summary-dense}
  \scriptsize
  \begin{tabular}{ccccc}
    \toprule
    Graph group & Restarts $k$ & QAOA mean & QAOA std. & QASPA fixed \\
    \midrule
    Backbone $p=0.5$ & 1 & 0.778 & 0.076 & 0.987 \\
Backbone $p=0.5$ & 2 & 0.800 & 0.050 & 0.987 \\
Backbone $p=0.5$ & 3 & 0.809 & 0.053 & 0.987 \\
Backbone $p=0.5$ & 4 & 0.813 & 0.054 & 0.987 \\
Backbone $p=0.5$ & 5 & 0.813 & 0.054 & 0.987 \\
Backbone $p=0.5$ & 6 & 0.813 & 0.054 & 0.987 \\
Backbone $p=0.5$ & 7 & 0.819 & 0.060 & 0.987 \\
Backbone $p=0.5$ & 8 & 0.819 & 0.060 & 0.987 \\
Backbone $p=0.5$ & 9 & 0.819 & 0.060 & 0.987 \\
Backbone $p=0.5$ & 10 & 0.826 & 0.055 & 0.987 \\
Backbone $p=0.7$ & 1 & 0.817 & 0.067 & 0.988 \\
Backbone $p=0.7$ & 2 & 0.846 & 0.039 & 0.988 \\
Backbone $p=0.7$ & 3 & 0.854 & 0.039 & 0.988 \\
Backbone $p=0.7$ & 4 & 0.854 & 0.039 & 0.988 \\
Backbone $p=0.7$ & 5 & 0.854 & 0.039 & 0.988 \\
Backbone $p=0.7$ & 6 & 0.854 & 0.039 & 0.988 \\
Backbone $p=0.7$ & 7 & 0.854 & 0.039 & 0.988 \\
Backbone $p=0.7$ & 8 & 0.854 & 0.039 & 0.988 \\
Backbone $p=0.7$ & 9 & 0.854 & 0.039 & 0.988 \\
Backbone $p=0.7$ & 10 & 0.854 & 0.039 & 0.988 \\
\bottomrule

  \end{tabular}
\end{table*}

Table~\ref{tab:benchmark-summary-dense} shows that increasing the
additional-edge probability does not qualitatively change this pattern in the
reported means.  QAOA expected ratios improve with restarts and start closer to
the fixed QASPA selection line at $p_{\mathrm{edge}}=0.7$, but the fixed QASPA
selection score remains higher within the ten-start budget used here.

\begin{table*}[t]
  \centering
  \caption{Experimental summary for the nineteen hand-specified custom topologies.}
  \label{tab:benchmark-summary-custom}
  \scriptsize
  \begin{tabular}{ccccc}
    \toprule
    Graph group & Restarts $k$ & QAOA mean & QAOA std. & QASPA fixed \\
    \midrule
    Custom topologies & 1 & 0.771 & 0.108 & 1.000 \\
Custom topologies & 2 & 0.794 & 0.072 & 1.000 \\
Custom topologies & 3 & 0.798 & 0.075 & 1.000 \\
Custom topologies & 4 & 0.798 & 0.075 & 1.000 \\
Custom topologies & 5 & 0.798 & 0.075 & 1.000 \\
Custom topologies & 6 & 0.798 & 0.075 & 1.000 \\
Custom topologies & 7 & 0.798 & 0.075 & 1.000 \\
Custom topologies & 8 & 0.798 & 0.075 & 1.000 \\
Custom topologies & 9 & 0.798 & 0.075 & 1.000 \\
Custom topologies & 10 & 0.798 & 0.075 & 1.000 \\
\bottomrule

  \end{tabular}
\end{table*}

The aggregate custom-topology row in Table~\ref{tab:benchmark-summary-custom}
does not show how individual graph shapes behave under QASPA readout.  Table
\ref{tab:topology-shot-readout} therefore separates the nineteen hand-specified
topologies and compares the exact QASPA selection score with the finite-shot
score obtained from Equation~\eqref{eq:shot-rule}.  This is not a new QAOA
comparison; it is a readout check for the QASPA classical selection stage across
named graph shapes.  Reporting each topology also prevents the aggregate value
from hiding failures on individual shapes and prepares the later edge-order
study, where noncommuting blocks are tied to concrete graph structures rather
than to an anonymous average.

\begin{table*}[t]
  \centering
  \caption{Topology-level QASPA readout sanity check.  The exact column uses the four
  largest state-vector probabilities.  The sampled columns repeat
  $N_{\mathrm{shots}}=1025$ draws $50$ times and then apply the same QASPA
  classical selection rule to the observed counts.}
  \label{tab:topology-shot-readout}
  \scriptsize
  \setlength{\tabcolsep}{3.2pt}
  \begin{tabular}{lccccccc}
    \toprule
    Topology & Edges & Optimum & Shots & Exact sel. & Sampled mean & Sampled std. & Opt. rate \\
    \midrule
    Path & 7 & 7 & 1025 & 1.000 & 1.000 & 0.000 & 1.00 \\
Cycle & 8 & 8 & 1025 & 1.000 & 1.000 & 0.000 & 1.00 \\
Star & 7 & 7 & 1025 & 1.000 & 1.000 & 0.000 & 1.00 \\
Wheel & 14 & 10 & 1025 & 1.000 & 1.000 & 0.000 & 1.00 \\
Ladder 2x4 & 10 & 10 & 1025 & 1.000 & 1.000 & 0.000 & 1.00 \\
Cube & 12 & 12 & 1025 & 1.000 & 1.000 & 0.000 & 1.00 \\
Prism-tail & 11 & 9 & 1025 & 1.000 & 1.000 & 0.000 & 1.00 \\
Lollipop & 10 & 8 & 1025 & 1.000 & 1.000 & 0.000 & 1.00 \\
Barbell & 9 & 7 & 1025 & 1.000 & 1.000 & 0.000 & 1.00 \\
House-tail & 9 & 8 & 1025 & 1.000 & 1.000 & 0.000 & 1.00 \\
Bowtie-tail & 9 & 7 & 1025 & 1.000 & 1.000 & 0.000 & 1.00 \\
Sparse bipartite & 10 & 10 & 1025 & 1.000 & 1.000 & 0.000 & 1.00 \\
Chordal cycle & 12 & 8 & 1025 & 1.000 & 1.000 & 0.000 & 1.00 \\
Dense core with leaves & 13 & 9 & 1025 & 1.000 & 1.000 & 0.000 & 1.00 \\
Two squares & 10 & 10 & 1025 & 1.000 & 1.000 & 0.000 & 1.00 \\
Fan & 13 & 10 & 1025 & 1.000 & 1.000 & 0.000 & 1.00 \\
Zigzag with chords & 13 & 10 & 1025 & 1.000 & 1.000 & 0.000 & 1.00 \\
Bridge mesh & 10 & 10 & 1025 & 1.000 & 1.000 & 0.000 & 1.00 \\
Ring with hubs & 13 & 10 & 1025 & 1.000 & 1.000 & 0.000 & 1.00 \\
\bottomrule

  \end{tabular}
\end{table*}

The topology-level check remains local to the retained custom set, rather than
serving as a broad sampling-robustness claim.  In this restricted set, repeated
finite readout matches the exact QASPA selection result for the nineteen graph
shapes: each sampled mean is $1.000$ with zero observed standard deviation over
the $50$ repeated readouts.  The result supports the stability of the prescribed
classical selection rule for these small retained topologies, while leaving
larger or more degenerate graph families to future finite-shot studies.

Because QASPA edge blocks on adjacent edges need not commute, the edge order is
part of the algorithmic specification.  Table~\ref{tab:edge-order-sensitivity}
checks how much the QASPA selection score changes when the canonical edge order
is replaced by the reverse order or by random permutations.  For each graph, the
random-order column uses $20$ sampled edge permutations; the table reports group
means and the range observed across those random-order trials.  The result is
not order-invariant, but the shuffled means stay close to the canonical
fixed-order values in these small tests.

\begin{table*}[t]
  \centering
  \caption{Sensitivity of the QASPA selection ratio to edge ordering.
  Random-order statistics use $20$ edge permutations per graph.}
  \label{tab:edge-order-sensitivity}
  \scriptsize
  \begin{tabular}{lccccc}
    \toprule
    Graph group & Canonical & Reversed & Random mean & Random std. & Random range \\
    \midrule
    Backbone $p=0.1$ & 0.990 & 1.000 & 0.999 & 0.011 & 0.875--1.000 \\
Backbone $p=0.3$ & 0.955 & 0.972 & 0.967 & 0.052 & 0.800--1.000 \\
Backbone $p=0.5$ & 0.987 & 0.958 & 0.970 & 0.051 & 0.714--1.000 \\
Backbone $p=0.7$ & 0.988 & 0.886 & 0.959 & 0.055 & 0.667--1.000 \\
Custom topologies & 1.000 & 0.978 & 0.991 & 0.030 & 0.800--1.000 \\
\bottomrule

  \end{tabular}
\end{table*}

Taken together, Figure~\ref{fig:restart-profile} and
Tables~\ref{tab:benchmark-summary}--\ref{tab:edge-order-sensitivity} support two
modest claims about optimization budget and shallow-depth baselines.  First,
QAOA's reported quality is sensitive to the classical optimization budget,
since one optimizer start is not the same experiment as ten starts.  Second,
the QASPA edge-encoded circuit places large-cut candidates inside the retained
candidate set without an outer-loop angle search in these tests.
The comparison is not a hardware or scaling claim, but it provides a controlled
indication that QASPA-like edge-encoded mixing is worth testing as a way to
change how much outer-loop tuning is needed.

\section{Discussion and Limitations}

The mathematical difference between the two approaches is where graph
structure enters the circuit.  In depth-one QAOA, the unweighted edge set
defines a diagonal phase separator and the global $X$ mixer then converts
phase differences into amplitude changes.  In \qaspa, each edge immediately
applies a conditional $Y$-axis rotation, so part of the edge structure is
embedded into the mixing step itself.  This resembles the broader trend in
multi-angle, free-axis, and custom-mixer QAOA variants: shallow circuits often
use more problem structure inside the ansatz when they are designed to reduce
the burden on depth or repeated classical tuning.

The comparison also exposes a trade-off that matters for later hardware-facing
work.  At the usual gate-decomposition level, both one-layer QASPA and
depth-one QAOA can require edge-local entangling structure: QASPA uses a
$\mathrm{CX}$--$R_y$--$\mathrm{CX}$ block per ordered edge, while QAOA cost
unitaries are commonly compiled through edge-local $ZZ$ rotations.  The
difference emphasized here is not a universal two-qubit-gate advantage, but the
placement of graph structure inside a noncommuting ordered QASPA mixing layer
instead of a diagonal QAOA phase layer followed by a global mixer.  This affects
scheduling, edge-order sensitivity, and the amount of classical parameter
search associated with obtaining large-cut samples.

The QASPA selection score is intentionally reported because it is the classical
optimization stage of the algorithm studied here, not a second search routine
applied after the state has been prepared.  The classical work after
measurement is the evaluation of four candidate bit strings by the original
Max-Cut objective, which is also the verification step needed to report any
sampled cut value.  This choice makes QASPA a candidate-generation method
rather than an expectation-value optimizer.  For that reason the paper also
reports the QASPA expectation value as a metric-control quantity and avoids
interpreting the selection-score comparison as a variational-optimization
dominance claim.

The limitations are important for reading the result at the right scale.  The
experiments use only $n=8$ exact state-vector simulations, so all results are
small-instance evidence.  The QAOA baseline is depth-one and uses a standard
local optimizer with a fixed iteration limit; the diagnostic table records
starts that terminate at that limit rather than hiding them.  Deeper QAOA,
warm-started QAOA, transferred parameters, CVaR objectives, or other optimizer
choices could change the comparison.  The QASPA edge order is also part of the
algorithmic definition.  The sensitivity check shows that random edge orders
give nearby group means in the tested cases, but it does not prove order
robustness for larger graphs.  The current QASPA construction is restricted to
unweighted graphs, since a weighted Max-Cut variant would need a separate
derivation rather than the direct substitution $\theta_{ij}=\alpha w_{ij}$.
The topology readout test samples repeatedly from an ideal state-vector
distribution, so a next version should add device-aware transpilation and noise
models.  It should also test larger unweighted graph families where exact
optima are replaced by independently verified classical baselines.

\section{Implementation Details and Reproducibility}

This section expands the implementation path behind the compact listings used
throughout the manuscript.  Its role is not to introduce a second algorithm, but
to make the computational workflow readable as a sequence of small, checkable
operations.  The same convention is used throughout: graphs are unweighted, bit
strings are q0-first, and the expected cut is computed exactly from the
state-vector probabilities when optimizing QAOA, while the main QASPA score
uses exact probabilities inside its classical selection stage.  The topology
readout test then repeats that QASPA selection step on sampled counts.

Figure~\ref{fig:repro-workflow} gives the section-level workflow before the
individual code fragments are introduced.  It connects graph generation, exact
state construction, QAOA restarts, sampled QASPA readout, and manuscript tables
inside one reproducibility path.

\noindent\begin{minipage}{\linewidth}
  \centering
  \includegraphics[width=\linewidth]{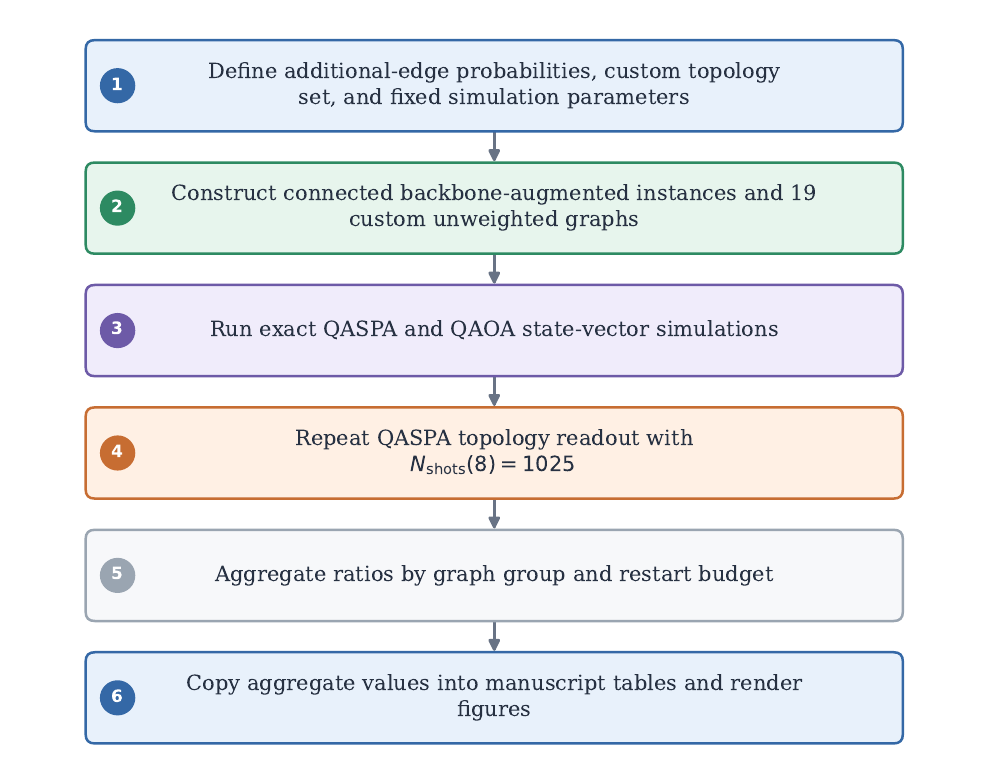}
  \captionof{figure}{Reproducibility workflow used for the figures and tables in the
  manuscript.}
  \label{fig:repro-workflow}
\end{minipage}

Listing~\ref{lst:graph-generator} shows the graph construction rule in
algorithmic form.  A path backbone is inserted first so that every generated
instance is connected, and additional edges are then added according to
$p_{\mathrm{edge}}$.  This makes the random graph family simple enough to
describe in the paper while avoiding isolated vertices that would not affect
the Max-Cut objective.

\begin{lstlisting}[style=paperpseudo,caption={Connected unweighted graph generation.},label={lst:graph-generator}]
def connected_random_graph(n, p_edge, seed):
    rng = random_generator(seed)
    edges = [(i, i + 1) for i in range(n - 1)]
    existing = set(edges)

    for i in range(n):
        for j in range(i + 1, n):
            if (i, j) not in existing and rng.random() < p_edge:
                edges.append((i, j))
                existing.add((i, j))

    if len(edges) == n - 1 and n > 2:
        edges.append((0, n - 1))

    return edges
\end{lstlisting}

The experiment also includes nineteen explicit graph topologies that are not
sampled from the additional-edge model.  Table~\ref{tab:custom-topologies} lists
these topologies immediately after the generator fragment, so the random and
hand-specified graph sources are introduced in one continuous methodological
block.

\begin{center}
\begin{minipage}{\linewidth}
  \centering
  \captionof{table}{Hand-specified unweighted custom graph topologies used as a separate
  experimental group.}
  \label{tab:custom-topologies}
  \scriptsize
  \setlength{\tabcolsep}{3.5pt}
  \begin{tabular}{@{}lcc@{}}
    \toprule
    Topology & Edges & Exact optimum \\
    \midrule
    Path & 7 & 7 \\
    Cycle & 8 & 8 \\
    Star & 7 & 7 \\
    Wheel & 14 & 10 \\
    Ladder 2x4 & 10 & 10 \\
    Cube & 12 & 12 \\
    Prism-tail & 11 & 9 \\
    Lollipop & 10 & 8 \\
    Barbell & 9 & 7 \\
    House-tail & 9 & 8 \\
    Bowtie-tail & 9 & 7 \\
    Sparse bipartite & 10 & 10 \\
    Chordal cycle & 12 & 8 \\
    Dense core with leaves & 13 & 9 \\
    Two squares & 10 & 10 \\
    Fan & 13 & 10 \\
    Zigzag with chords & 13 & 10 \\
    Bridge mesh & 10 & 10 \\
    Ring with hubs & 13 & 10 \\
    \bottomrule
  \end{tabular}
\end{minipage}
\end{center}

The custom list makes the nonrandom part of the experiment auditable.  The
topologies cover sparse trees, cycles, bipartite graphs, dense cores with
leaves, and mixed bridge structures while preserving the unweighted-graph
restriction of the QASPA construction.

The QASPA state update is deterministic once the edge order and angle are
fixed.  The depth parameter is retained in Listing~\ref{lst:qaspa-full} to
make clear how a later experiment could repeat the same edge schedule, even
though the numerical comparison in this manuscript uses one layer.

\begin{lstlisting}[style=paperpseudo,caption={Full QASPA state construction used in the experiment.},label={lst:qaspa-full}]
def qaspa_state(edges, n, alpha, depth=1):
    state = zero_state(2**n)
    state[0] = 1

    for q in range(n):
        apply_h(state, q)

    for layer in range(depth):
        for i, j in edges:
            apply_cx(state, i, j)
            apply_ry(state, j, alpha)
            apply_cx(state, i, j)

    return state
\end{lstlisting}

For the QAOA baseline, the state starts from $|+\rangle^{\otimes n}$, receives
the diagonal phase generated by the Max-Cut cost, and then receives the global
$X$ mixer.  Listing~\ref{lst:qaoa-state} follows
Equations~\eqref{eq:qaoa-state} and~\eqref{eq:qaoa-unitaries} closely, making
the transition from the mathematical expression to the exact state-vector
calculation explicit.

\listingneedspace{18\baselineskip}
\begin{lstlisting}[style=paperpseudo,caption={Depth-one QAOA state construction.},label={lst:qaoa-state}]
def qaoa_state(edges, n, gamma, beta):
    state = plus_state(n)
    cut_values = [cut_value(mask, edges) for mask in range(2**n)]

    for mask in range(2**n):
        state[mask] *= exp(-1j * gamma * cut_values[mask])

    for q in range(n):
        apply_rx_mixer(state, q, beta)

    return state
\end{lstlisting}

The restart comparison is a classical wrapper around this state preparation.
A single local optimization can stop in a poor region of the two-dimensional
parameter landscape, so the restart routine records the best value after the
first start, after the first two starts, and so on until ten starts.  This
best-so-far trace is the quantity shown in Figure~\ref{fig:restart-profile} and
Tables~\ref{tab:benchmark-summary}--\ref{tab:benchmark-summary-custom}, where the
sparse random, denser random, and hand-specified topology groups are reported
separately.
Listing~\ref{lst:best-of-k} gives this best-of-$k$ wrapper explicitly.

\begin{lstlisting}[style=paperpseudo,caption={Best-of-k QAOA restart wrapper.},label={lst:best-of-k}]
def qaoa_best_of_k(edges, n, k_max):
    best = -infinity
    trace = []

    for k in range(1, k_max + 1):
        gamma0, beta0 = random_angles()
        gamma, beta = local_search(edges, n, gamma0, beta0)
        state = qaoa_state(edges, n, gamma, beta)
        ratio = expected_cut(state, edges) / exact_optimum(edges, n)
        best = max(best, ratio)
        trace.append((k, best))

    return trace
\end{lstlisting}

The topology readout test uses the same QASPA classical selection rule, but
the four candidates are obtained from observed counts rather than from exact
probability ordering.  Listing~\ref{lst:qaspa-shot-readout} records this
finite-shot wrapper separately so that it is not confused with the QAOA
restart loop.

\begin{lstlisting}[style=paperpseudo,caption={Finite-shot QASPA classical selection for topology tests.},label={lst:qaspa-shot-readout}]
def qaspa_sampled_selection(state, graph, n):
    shots = ceil(0.03047 * 2**n + 1016.2)
    counts = sample(abs(state)**2, shots)

    selected = four_most_frequent_bitstrings(counts)
    best_cut = max(cut_value(mask, graph.edges) for mask in selected)
    return best_cut / exact_optimum(graph)
\end{lstlisting}

The exact optimum is computed by exhaustive enumeration in these small
instances, which is not a scalable Max-Cut solver but is appropriate for $n=8$
because it gives an unambiguous denominator for the approximation ratio.  The
enumeration routine is shown in Listing~\ref{lst:exact-optimum}.

\begin{lstlisting}[style=paperpseudo,caption={Exact optimum for the small graphs in this manuscript.},label={lst:exact-optimum}]
def exact_optimum(edges, n):
    best_value = -infinity
    best_masks = []

    for mask in range(2**n):
        value = cut_value(mask, edges)
        if value > best_value:
            best_value = value
            best_masks = [mask]
        elif value == best_value:
            best_masks.append(mask)

    return best_value, best_masks
\end{lstlisting}

\section{Future Work and Scope}

The main experimental result is reported as a mean over graph groups rather than
as a claim about every graph.  This matters because
variational optimization can behave unevenly across instances.  Some instances
can be easy for one ansatz and hard for another, and small changes in the
optimizer starting point can change the local optimum found by QAOA.  The
restart curve is a diagnostic of classical search sensitivity, not a
universal ordering of algorithms.

The fixed QASPA line functions as a low-classical-search baseline.  It does not
require a two-dimensional optimizer, and in the fixed-angle condition it does
not require any instance-specific angle selection.  This keeps the
contrast with QAOA focused on the visible restart budget rather than on two
different forms of classical parameter search.

The QAOA line is intentionally shown as best-of-$k$ rather than as ten
independent unrelated numbers.  A practitioner who can afford ten local
starts will keep the best result, so the best-of-$k$ curve is the relevant
budget curve.  The plateau in this experiment means that, after several
starts, additional random initializations did not substantially change the
mean expected ratio under the chosen optimizer and graph group.  It does not
mean that no other optimizer, initialization rule, or QAOA depth could improve
the result.

The edge-block view also clarifies why unweighted graphs are the scope adopted
in the current paper.  The QASPA circuit uses the same angle on each edge,
and the graph structure enters through the placement and ordering of edge
blocks.  Simply multiplying the angle by a weight would change the
interpretation of the block and would need to be justified as a separate
algorithm.  The present manuscript keeps the scientific question narrow: what
can be learned from edge-encoded $R_y$ mixing on unweighted Max-Cut instances?

Figure~\ref{fig:future-workflow} summarizes the natural extension path from the
present exact simulations toward more demanding evaluations.

\begin{center}
  \begin{minipage}{\linewidth}
    \centering
    \includegraphics[width=\linewidth]{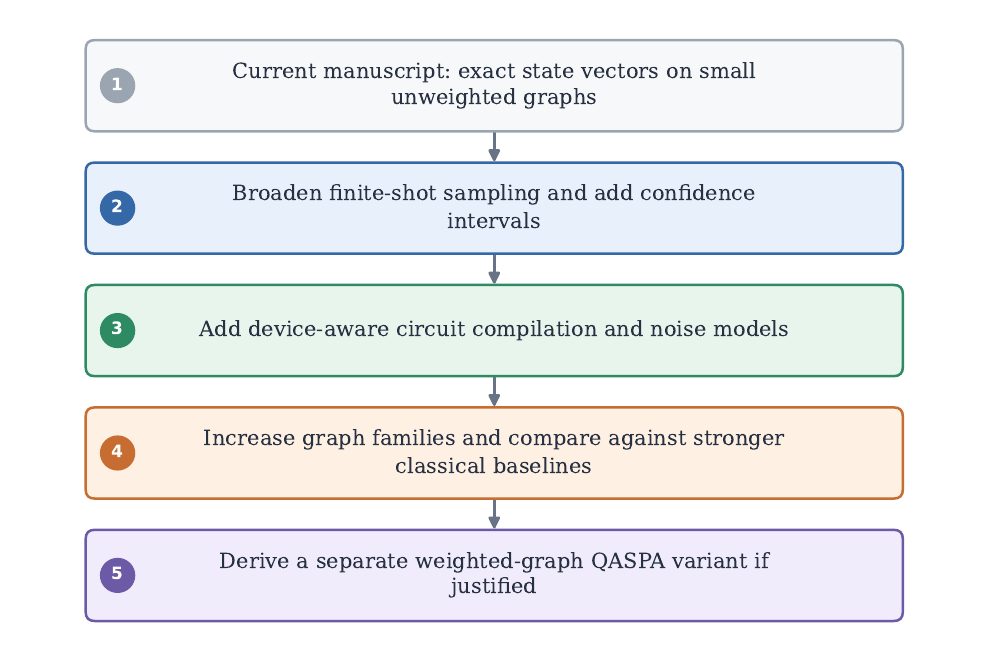}
    \captionof{figure}{Suggested extension path beyond the present exact-simulation study,
    moving from broader finite-shot studies toward a separately derived weighted variant.}
    \label{fig:future-workflow}
  \end{minipage}
\end{center}

The next step is to add one complication at a time: broader finite-shot
uncertainty intervals, hardware-aware compilation, larger unweighted graph
families, and then a separately derived weighted variant if the unweighted
behavior remains promising.  This sequence helps prevent the present result from
being overread as a hardware or scaling claim.  The experiments in this paper
test whether the fixed QASPA state places large cuts inside the candidate set
chosen by its classical selection stage on small unweighted graphs, and they
compare that behavior with the restart sensitivity of depth-one QAOA.  They do
not estimate device noise, routing overheads, asymptotic scaling, or the effect
of stronger classical baselines.  Those questions are best added after the
finite-shot behavior is measured across a broader collection of unweighted
instances, because otherwise sampling variation, optimizer sensitivity, and
hardware effects would be mixed into one ambiguous performance number.

The order in Figure~\ref{fig:future-workflow} is deliberately conservative.
Finite-shot uncertainty is still placed before hardware mapping because sampling
variation is easier to interpret when it is separated from device noise.
Stronger classical baselines are placed before any weighted-graph extension
because the present QASPA block does not define a weight-dependent angle rule.
This ordering keeps future work connected to the assumptions used in the
current experiments.

Table~\ref{tab:future-checks} lists the checks that should be added before
moving from the present paper to a hardware-facing claim, and it records the
boundary between this manuscript and the next research stage rather than
introducing a new experiment.

\begin{center}
  \scriptsize
  \captionof{table}{Future checks that should precede any hardware-facing claims about
  QASPA performance.}
  \label{tab:future-checks}
  \begin{tabular}{lll}
    \toprule
    Check & Why it matters & Current status \\
    \midrule
    Finite-shot study & Sampling variance & $50$ readouts \\
    Noise model & Gate errors & Not modeled \\
    Mapping & 2-qubit routing & Not modeled \\
    Larger $n$ & Scaling & Not included \\
    Baselines & Separation & Future work \\
    Weighted variant & Angle rule & Future work \\
    \bottomrule
  \end{tabular}
\end{center}

\section{Conclusion}

This paper develops \qaspa\ as a reproducible Max-Cut study by defining the
edge-encoded $R_y$ block algebraically and showing the complete transformation
sequence on an unweighted test graph.  It also includes the graph, probability,
and schematic figures used in the text, and it adds a comparison against
depth-one QAOA under a visible restart budget.  The topology-level readout check
further shows how the proposed shot rule interacts with the QASPA classical
selection stage on hand-specified graph shapes.

The main empirical observation is that the fixed QASPA state places large-cut
strings among its highest-probability candidates in the tested instances, while
the QAOA expected ratio improves as additional classical optimizer starts are
allowed.  The next scientific question is whether this candidate-quality
behavior persists for larger unweighted graphs and noisy hardware-aware
execution.

\section*{Author Contribution Statement}

Dmytro Sapozhnyk conceived the study, developed the QASPA-MaxCut formulation,
implemented the simulations, designed the experiments, interpreted the
results, and wrote and revised the manuscript.  AI-assisted tools were used
only to generate schematic diagrams used as visual aids in the manuscript.  The
author reviewed all generated schemes and takes full responsibility for the
content of the work.

\section*{Data and Code Availability}

The numerical values reported in the paper are given directly in the tables and
figures, and the experimental protocol is specified by the graph families,
fixed seed, shot rule, QASPA classical selection rule, and QAOA restart budget
described in the main text.  A frozen supplementary source archive containing
the simulation code, raw numerical arrays, plotted data, and figure-generation
inputs is prepared for anonymous editor and referee inspection.  No external
datasets are required.  Upon acceptance, the public archive is intended to be
released under a FOSS license compatible with the journal's open-source policy.

\bibliographystyle{quantum}
\begingroup
\scriptsize
\setlength{\itemsep}{0pt}
\setlength{\parskip}{0pt}

\endgroup

\end{document}